\documentclass{PoS}

\newcommand{\nn}{\nonumber\\}
\newcommand{\beqa}{\begin{eqnarray}}
\newcommand{\eeqa}{\end{eqnarray}}
\newcommand{\be}{\begin{equation}}
\newcommand{\ee}{\end{equation}}

\title{Dileptons and Photons  at RHIC}

\ShortTitle{Dileptons and Photons at RHIC}

\author{\speaker{Ismail Zahed}\\
        Department of Physics and Astronomy\\
        SUNY Stony Brook, Stony Brook NY 11794\\
        E-mail: \email{zahed@tonic.physics.sunysb.edu}}

\abstract{Electromagnetic emission in the form of photons or dileptons provide
important information on the onset and evolution of a heavy ion collision
at ultrarelativistic energies. We briefly summarize the theoretical 
assessments of the thermalized electromagnetic emissivities from a hot
partonic and hadronic medium, and compare them to current experiments at 
collider energies.}

\FullConference{Light Cone 2010 - LC2010\\
		June 14-18, 2010\\
		Valencia, Spain}

\begin{document}

\section{Introduction}
At low and intermediate masses, the electromagnetic emissivities are dominated
by thermal emissions. Thermalization at RHIC has been established using both
statistical analysis (chemical equilibration)~\cite{STAT} and flow measurements 
(thermal equilibration)~\cite{FLOW}. PHENIX has reported dilepton
emissivities all the way to the Dalitz region~\cite{EE}, and more recently
extrapolated photon rates~\cite{EE}. In so far, the electron emissivities in the 
intermediate  mass region have defied most theoretical understanding.

In this talk, I will give a brief overview of our current theoretical understanding of the 
electromagnetic emissivities by focusing on the photon rates. I will quote the electronic
rates. In section 2, the hadronic photon rates are worked out in terms of the chiral
reduction formulae tying the rate to vacuum correlation functions. The hadronic rates 
overun the partonic rates for a broad range of photon energies. In section 3, we rely 
on a hydrodynamical  evolution to assess the electromagnetic emissivities for 3
distinct experiments to underline the consistency of our approach. Our summary
is in section 4.

\section{Hadronic Photon Rates}

For a hadronic gas in thermal equilibrium the number of photon produced per unit four 
volume and unit three momentum is tied to the electromagnetic current-current 
correlation function~\cite{Steele:1996su}

\beqa
\frac{q^0 dN}{d^3q}=-\frac{\alpha_{em}}{4\pi^2}\,\frac{2}{1+e^{q_0/T}}{\rm Im}\,{\bf W}^F(q)\,,
\eeqa
with $q^2=0$ and 

\beqa
{\bf W}^F(q)={\bf W}_0+\int d\pi_1 {\bf W}_\pi + \frac{1}{2!}\int d\pi_1 d\pi_2 {\bf W}_{\pi\pi}+\cdots
\eeqa
The expansion in ${\bf W}^F$ is carried over stable hadronic states. At RHIC the baryonic potential
is small, so the nucleons can be ignored in the expansion. Here

\beqa
d\pi_i = \frac{d^3k_i}{(2\pi)^3}\frac{n(E_i)}{2E_i}\,\,.
\eeqa
counts the pion phase space. We have defined 
\beqa
{\bf W}_0&=&i\int d^4x e^{iq\cdot x}\langle 0\vert T^* {\bf J}^\mu(x) {\bf J}_\mu(0)\vert 0\rangle \nn
{\bf W}_\pi &=&i\int d^4x e^{iq\cdot x}\langle \pi^a(k_1)\vert T^* {\bf J}^\mu(x) {\bf J}_\mu(0)\vert \pi^a(k_1)\rangle \nn
{\bf W}_{\pi\pi}&=&i\int d^4x e^{iq\cdot x}\langle \pi^a(k_1)\pi^b(k_2)\vert T^* {\bf J}^\mu(x) {\bf J}_\mu(0)\vert \pi^a(k_1)\pi^b(k_2)\rangle 
\label{WWW}
\eeqa
with the sum over isospin subsumed.  

The first contribution in (\ref{WWW}) is dominated by the transverse part of the isovector correlator and is entirely fixed experimentally by the measured electroproduction data.  It vanishes for real photons, i.e. ${\bf W}_0=0$.
The next two terms, ${\bf W}_\pi$ and ${\bf W}_{\pi\pi}$, can be reduced to measurable vacuum correlators by the the
chiral reduction formulae~\cite{CRF}.  For instance~\cite{Steele:1996su,CRF}

\beqa
{\bf W}^F_\pi(q,k)&=&\frac{12}{f_\pi^2}q^2\,{\rm Im} {\Pi}_V(q^2)\nn
&-&\frac{6}{f_\pi^2}(k+q)^2\,{\rm Im} { \Pi}_A \left( (k+q)^2\right) + (q\to -q)\nn
&+&\frac{8}{f_\pi^2}\left( (k\cdot q)^2-m_\pi^2 q^2\right) \,{\rm Im} { \Pi}_V(q^2)\times\,{\rm Re} \Delta_R(k+q)+(q\to-q)\nn
\label{eq:lin_in_meson1}
\eeqa
where ${\rm Re}\Delta_R$ is the real part of the retarded pion propagator, 
and ${\Pi}_V$ and ${\Pi}_A$ are the transverse parts of the VV and AA 
correlators.  Their spectral functions are related to both $e^+e^-$ annihilation 
and $\tau$-decay data. The two-pion reduced contribution ${\bf W}_{\pi\pi}$ is more 
involved~\cite{Steele:1996su,myThesis}.  

The dielectron rates follow exactly the same analysis since they correspond to
virtual photon emissivities corrected by leptonic matrix elements. Specifically,
for $q^2<0$,

\begin{equation}
\label{ee}
\frac{dR}{d^4q}=\frac{-\alpha^2}{3\pi^3 q^2}
\,\left(1+\frac{2m^2_l}{q^2}\right)\left(1-\frac{4m^2_l}{q^2}\right)^{1/2}
\frac{1}{1+e^{ q^0/T}}\,{\rm Im}{\bf W}^F(q)
\end{equation}
Thus

\begin{equation}
\frac{dR}{d^4q}=\frac{2\alpha}{3\pi\,q^2}
\,\left(1+\frac{2m^2_l}{q^2}\right)\left(1-\frac{4m^2_l}{q^2}\right)^{1/2}
\,\left(\frac {q^0\,dN^*}{d^3q}\right)
\label{EXTRA}
\end{equation}
which ties the dielectron rate to the {\it virtual} photon rate $N^*$  again for
spacelike momenta.  The chiral reduction approach to both dielectrons and
photons preserve the nature of this relation. This is not the case for most 
approaches based on specific hadronic processes.  This point is important
while analyzing both the electron and photon data reported by PHENIX,
since the latter are extrapolated from the former.

\begin{figure}[hbtp]
\includegraphics[scale=.5]{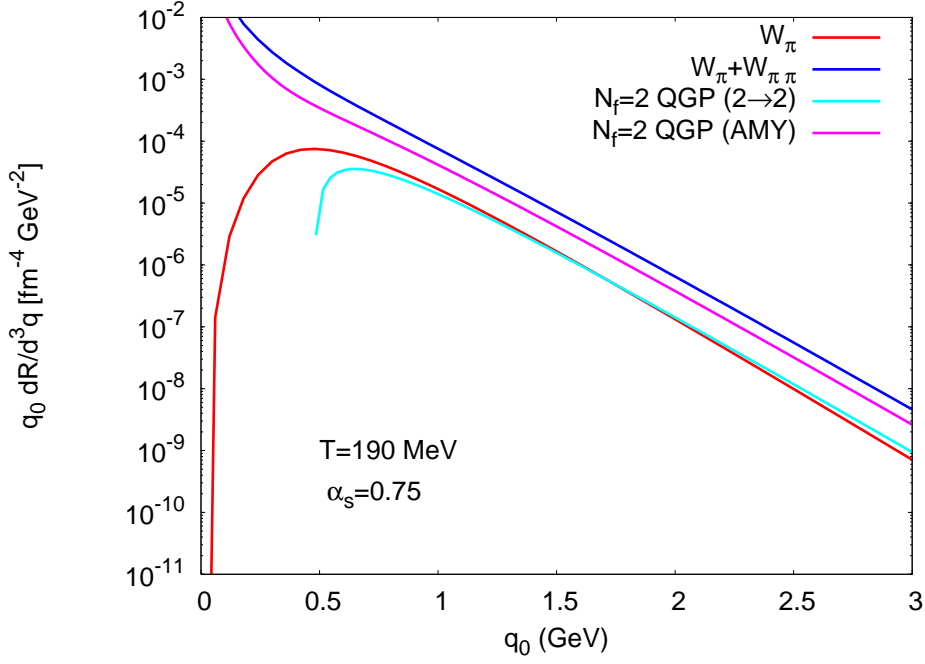}
\caption{Hadronic photon rates versus partonic rates.}
\label{fig:compareQGP}
\end{figure}

In Fig.~\ref{fig:compareQGP} we compare our hadronic rates with the complete leading order 
quark-gluon plasma rates~\cite{AMMY} for a broad range of photon energies  at the highest hadronic temperature
$T=190$ MeV  and for  a substantial coupling $\alpha_s=3/4$. We note that our full hadronic rates 
are consistently higher than the leading order QGP rates even at the highest photon energies. 
The hadronic bremsstralung at low photon energies dwarf the near-collinear bremsstralung from
the quarks and antiquarks in the QGP~\cite{DUSLINGZAHED}.

\section{Spectra in Ultra-relativistic Heavy-Ion Collisions}

The hadronic and partonic rates are usually convoluted with the time-evolution history
of the fireball from inception at about 1/2 fm/c to thermal freezeout at about 10 to 15 fm/c,
depending on the collider energy. The final evolved rates are then folded with the detector
acceptance to compare with experiments.  In table~\ref{tab:param} we summarize our 
initial parameters based on hydrodynamical evolution as in~\cite{DUSLINGZAHED}.
The equation of state used for the SPS evolution is based on the bag model, while the 
equation of state used for both RHIC and LHC is based on lattice results. The SPS 
evolution ignores baryons   in line with the hadronic expansion above. This approximation
fares poorly at the SPS, so our emissivity quotes should be viewed as lower bounds.

\begin{table}[h]
\begin{tabular}{l c c c c}
\hline
Parameter & SPS & RHIC 1 & RHIC 2 & LHC \\
\hline\hline
$\sqrt{s_{NN}}$ \mbox{ [A--GeV]} & 17.3 & 200 & 200 & 5500 \\
A & 208 & 197 & 197 & 208 \\
$\sigma_{NN}^{\mbox{in.}} \mbox{ [mb]}$ & 33 & 40 & 40 & 60 \\  
$C_s$ & 8.06 & 20.8 & 20.8 & 42 \\
$C_B$ & 0.191 & 0. & 0. & 0. \\
\hline
\mbox{  Centrality:} & 0-10\% & 0-20\% & 0-20\%  & 0-20\% \\
b [fm] & 3 & 4.5 & 4.5 & 4.8 \\
$N_{\mbox{part}}$ & 340 & 269 & 269 & 293 \\
$\tau_0$ \mbox{ [fm/c]}& 1 & 1 & 0.5 & 0.5 \\
$T({\bf r}_\perp=0,\tau_0) \mbox{ [MeV]}$ & 245 & 336 & 398 & 501 \\
$T_{frzout} \mbox{ [MeV]}$ & 120 & 140 & 160 & 140\\
\hline
\end{tabular}
\caption{Hydrodynamical parameters for: SPS, RHIC and LHC.}
\label{tab:param}
\end{table}

\subsection{SPS}

\begin{figure}[hbtp]
  \vspace{9pt}
  \centerline{\hbox{ \hspace{0.0in}
\includegraphics[scale=.35]{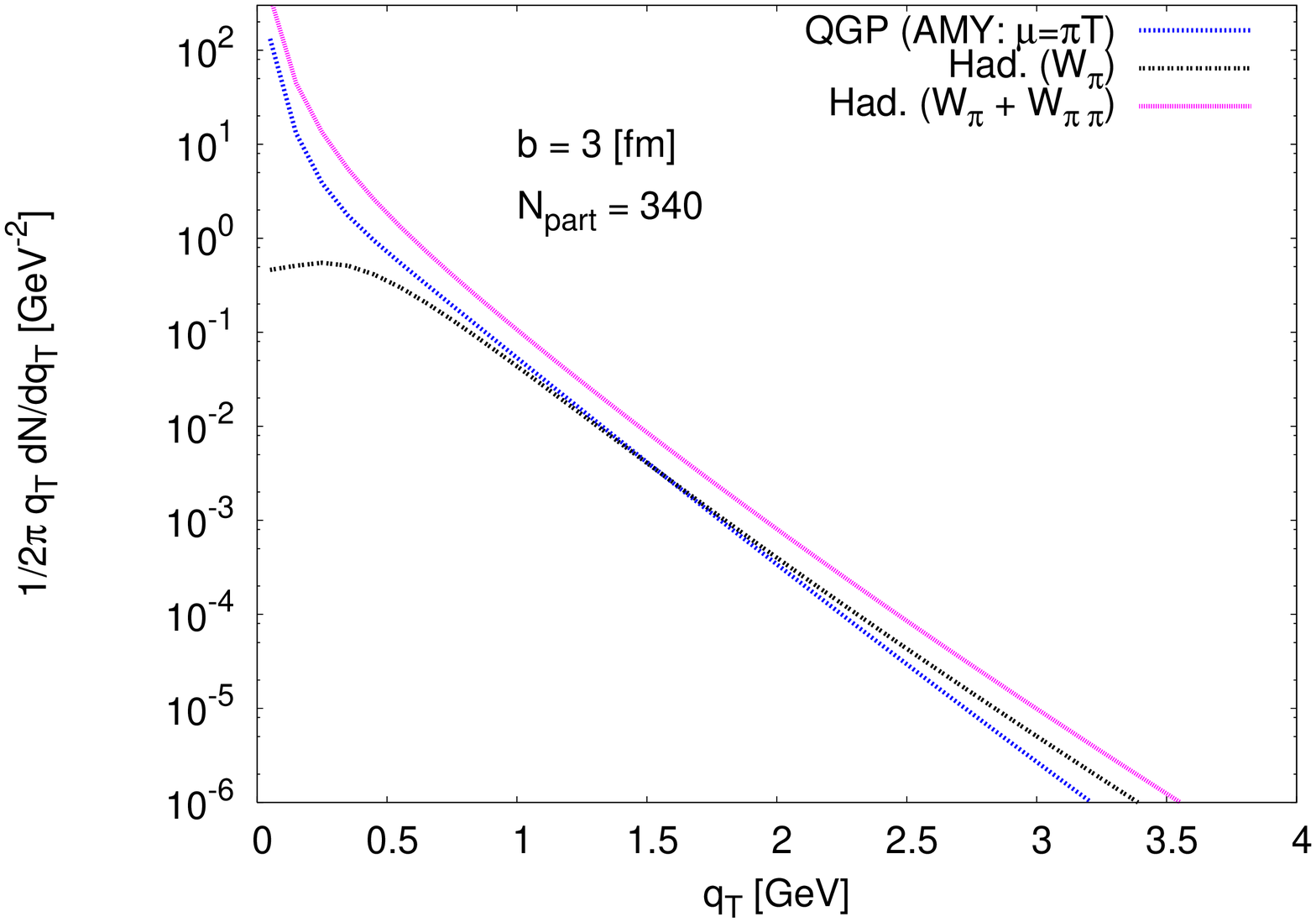}
    \hspace{0.0in}
\includegraphics[scale=.35]{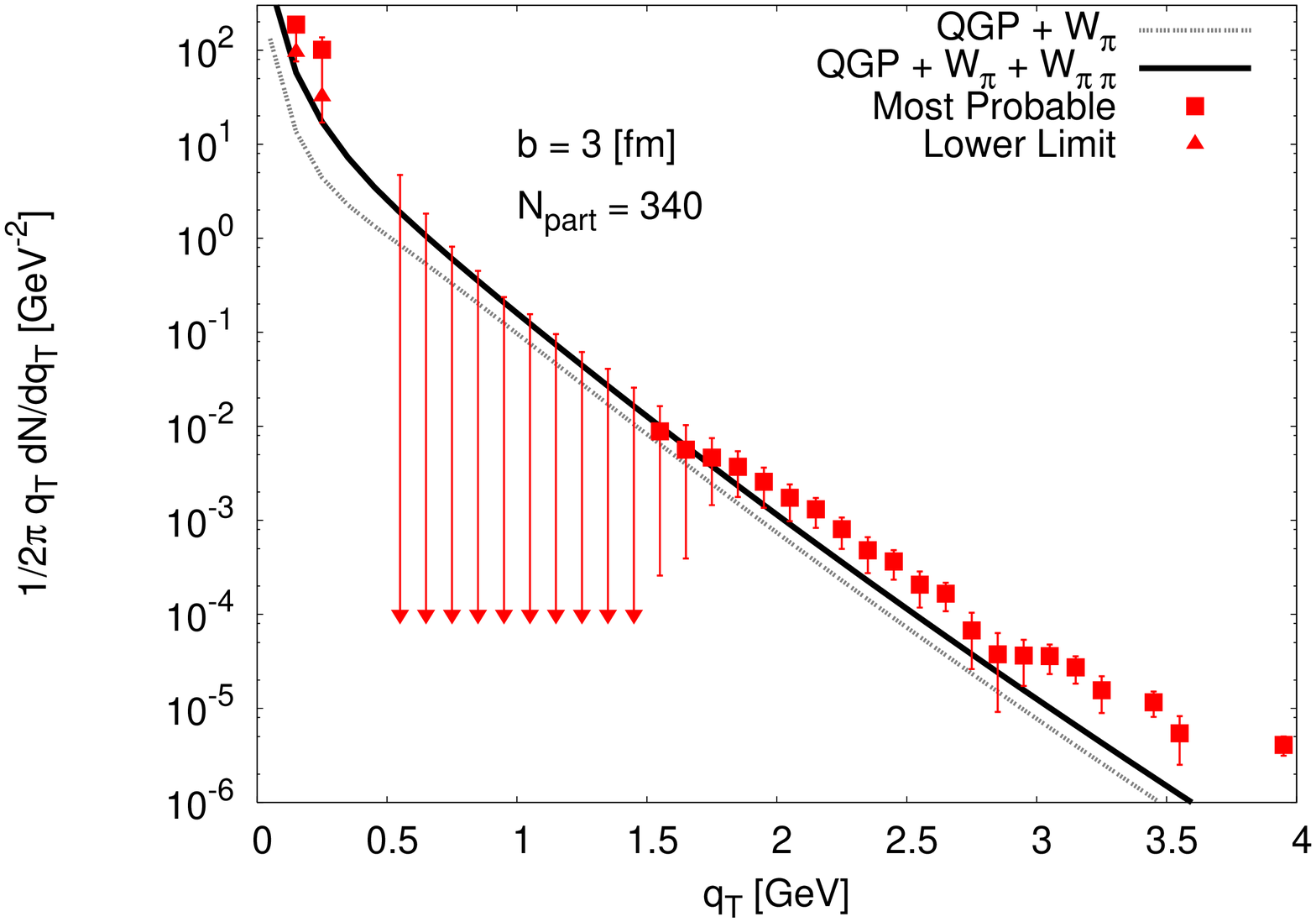}
    }
  }
  \caption{Photon Spectra at the SPS: Rate comparison (left) and Data comparison (right).}
  \label{figX}
\end{figure}

In Fig.~\ref{figX} (left) we detail our evolved photon emissivities from $W_\pi$ and $W_{\pi\pi}$ versus the improved
and leading order QGP rates for WA98  at an impact parameter of $b=3$ fm which involves about 340 participants
initially. The hadonic rates are substantially larger than the QGP rates both at low and high photon energies. In 
Fig.~\ref{figX} (right) we compare our rates with the data from~\cite{SPSDATA}. Remarkably, our hadronic rates 
fit the photon data both at very low and very high energy although on the lower side. Since our current rates do not
include baryons, this is expected. The inclusion of the latters should improve the fits both at low and high energy.  
In the intermediate mass region, our rates favor the current upper bounds from the data and is consistent with a
previous analysis using $W_\pi$ and the baryon contributions~\cite{Steele:1996su}.

\subsection{RHIC}

\begin{figure}
\includegraphics[scale=.32]{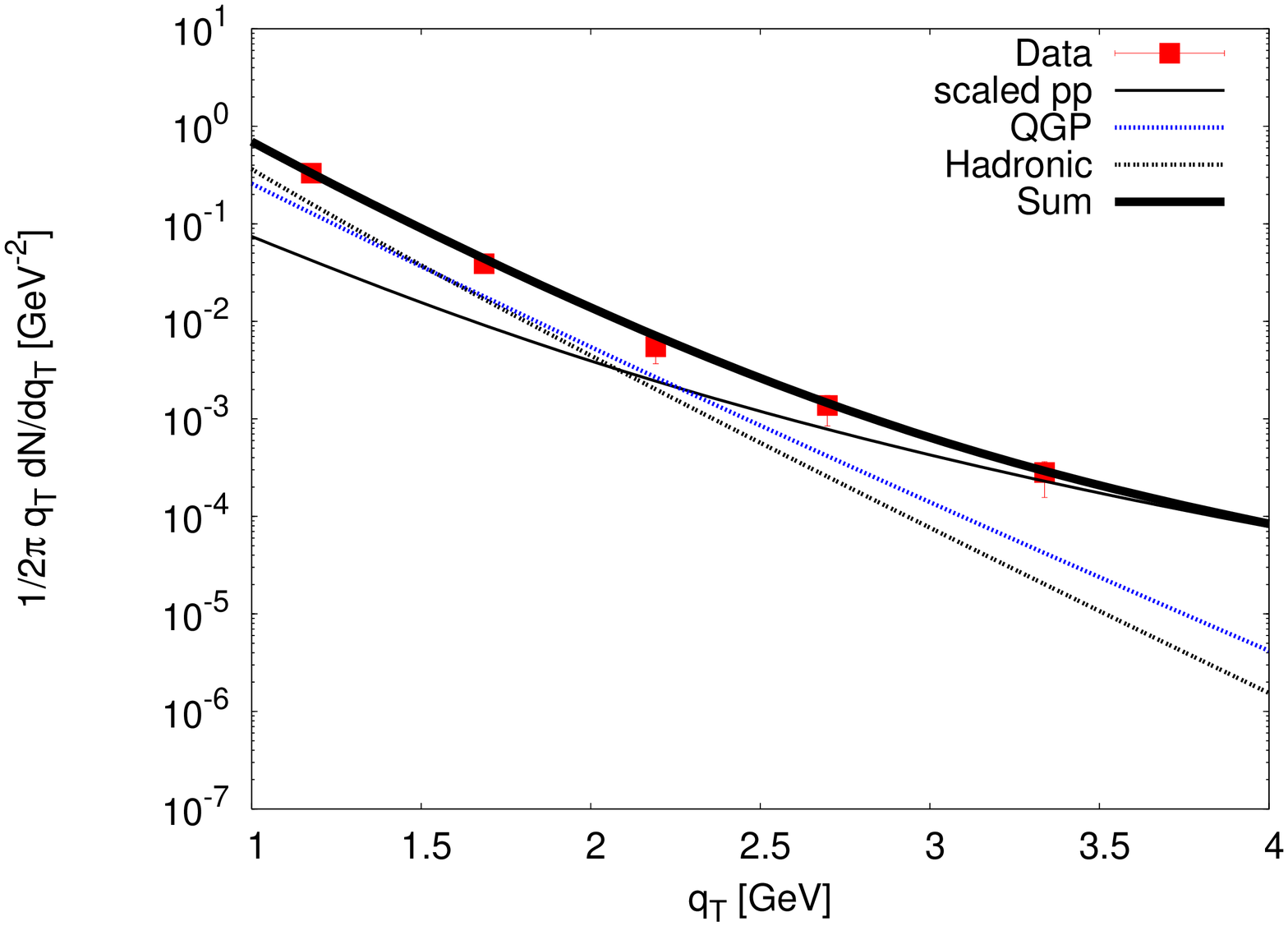}
\includegraphics[scale=.32]{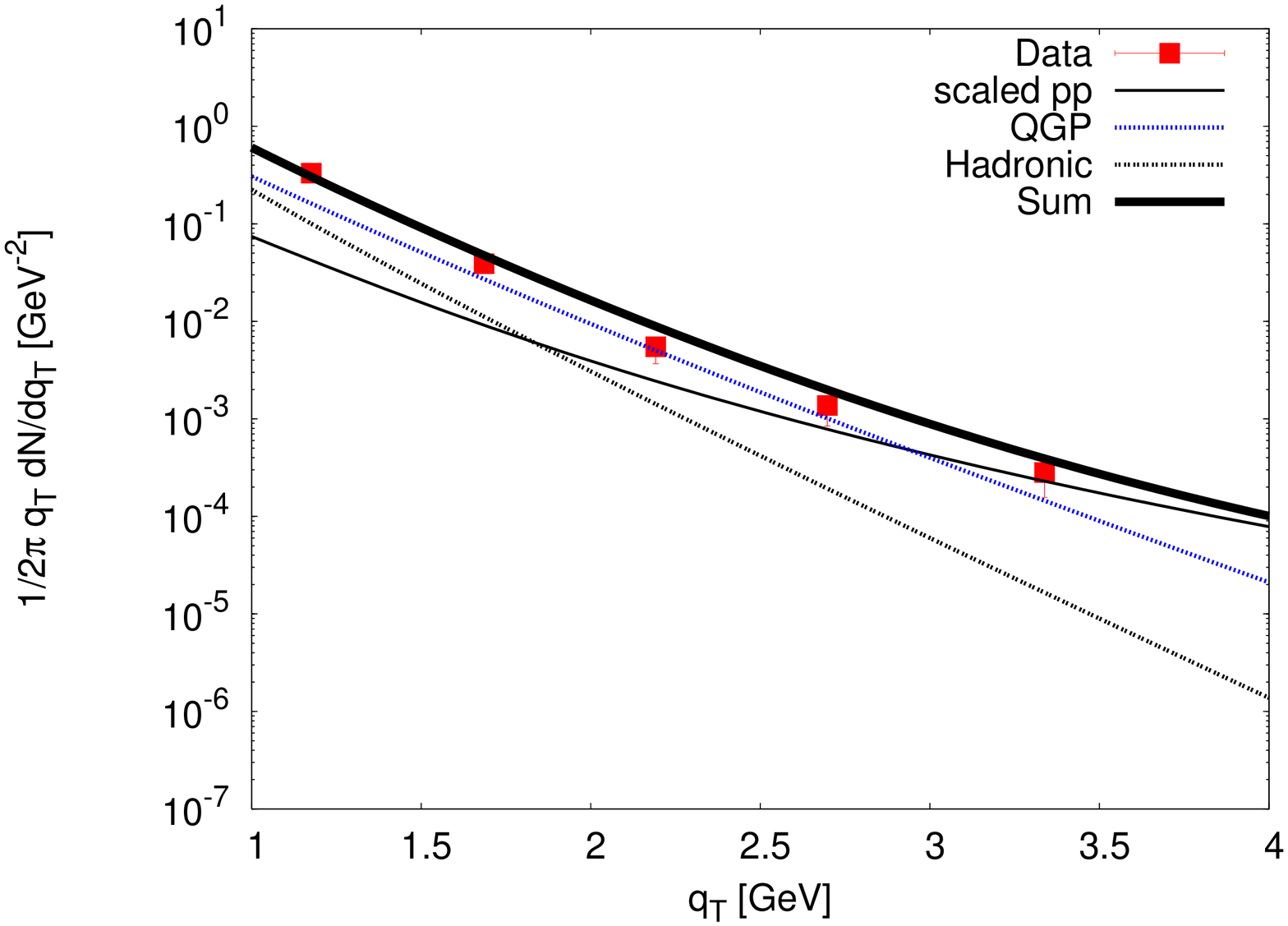}
\caption{Photon spectra at RHIC versus data: RHIC1 (left) and RHIC2 (right).}
\label{RHICDATA}
\end{figure}

In Fig.~\ref{RHICDATA} we show the evolved photon rates versus RHIC data for the two different
hydrodynamical set ups in Table~\ref{tab:param},  (left) is RHIC1 and (right) is RHIC2. RHIC2 is a
bit more {\it explosive} than RHIC1 although both account correctly for the final hadron multiplicities.
The scaled pp data by the number of participants are added to account for the contribution from the 
prompt pp collisions not included in our rate analysis. For RHIC1 our analysis is consistent with the
current data as reported above 1GeV. Interestingly enough, for RHIC1 the first two photon points 
around 1 GeV are well described by the addition of the hadronic rates. The scaled pp emissivities 
rapidly take over the QGP emissivities around 1.5 GeV to reproduce the data. The photon data
appears to favor the parameter set RHIC1 as ooposed to RHIC2. The latter is more explosive with 
a longer lifetime for the QGP phase than in RHIC1.

\subsection{Photon spectra at LHC}

\begin{figure}
\includegraphics[scale=.5]{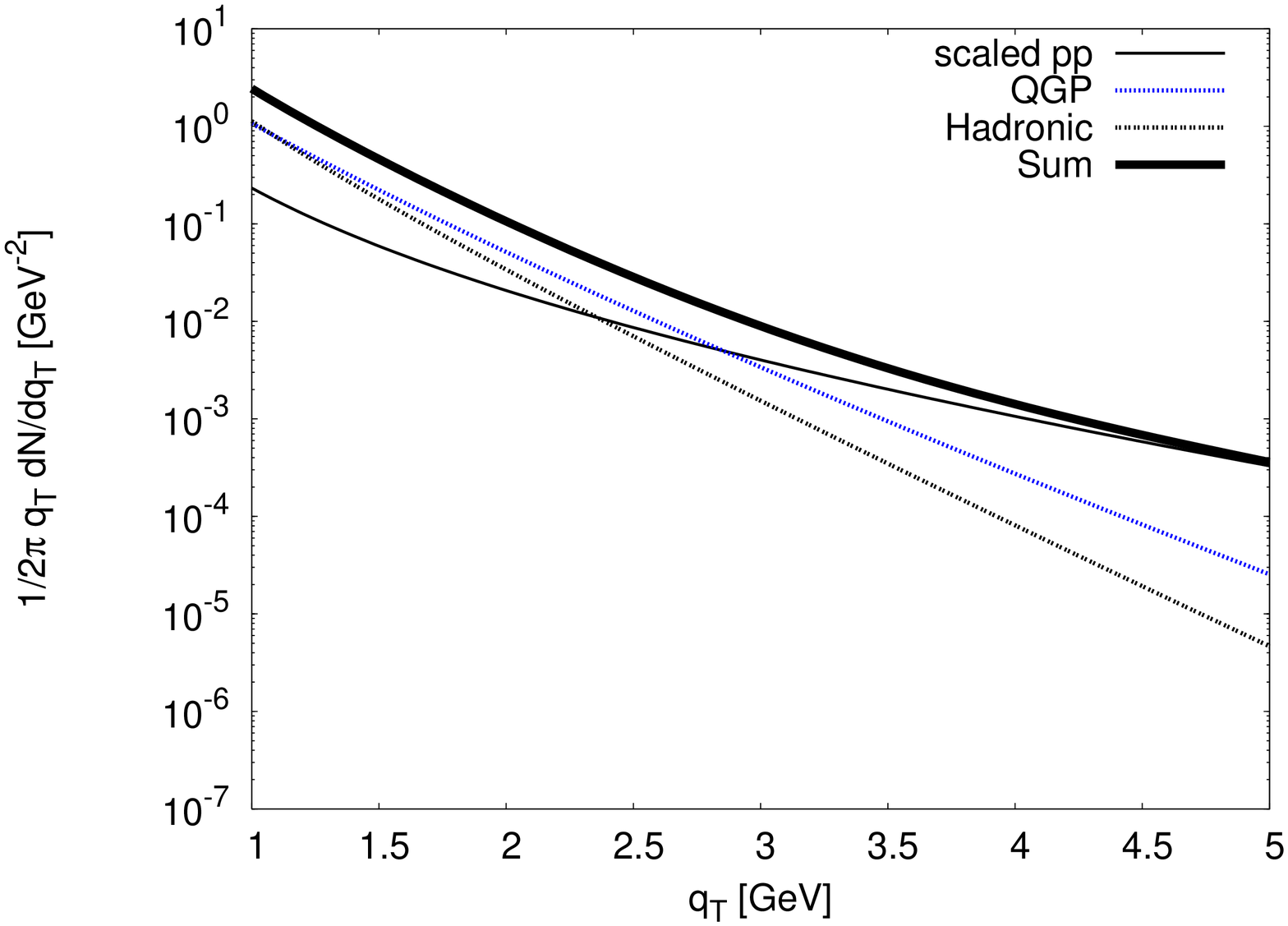}
\caption{LHC.}
\label{fig:lhc}
\end{figure}

In Fig.~\ref{fig:lhc} we display our projected photon emissivities at LHC using the 
hydrodynamical parameter set in Table~\ref{tab:param}.  We expect the conditions
at LHC to be a bit more explosive than RHIC, and therefore favoring a parameter
choice more in line with RHIC2.  The photon emissivities are QGP dominated between 
$1-2.5$ GeV with the hadronic emissivities being comparable to the QGP ones at about
1 GeV. The prompt pp photon emissivities take over the yield at about 2.5 GeV. 

\section{Conclusions}

Current theoretical approaches to the electromagnetic emissivities in ultrarelativistic heavy ion
colliders are well constrained by partonic/hadronic calculations within a broad range of energies.
The exception is the dilepton spectrum reported by PHENIX in the intermediate mass range, where
experiment exceeds the best calculations by almost a factor of 1.5. It is worth pointing that the 
extrapolated photon measurements by PHENIX are consistent with current calculations, making the
low mass dielectron discrepancy more puzzling. This said, it is clear that the strength and character
of the electromagnetic emissivities point to the formation of a primordial quark and gluon in heavy-ion 
collisions, with temperatures of the order of 350 MeV at RHIC and perhaps 500 MeV at the LHC.

\section{Acknowledgements}

The work presented in this talk was carried in collaboration with 
Kevin Dusling. This work was supported in part by US DOE grants DE-FG02-88ER40388
and DE-FG03-97ER4014.


\end{document}